\RequirePackage{fix-cm}
\documentclass[smallextended]{svjour3}       
\smartqed  
\usepackage{graphicx}
\usepackage{amssymb}
\usepackage{mathptmx}      
\usepackage{latexsym}
\journalname{International Journal for Philosophy of Religion}
\begin{document}

\title{New remarks on the Cosmological Argument}

\author{Gustavo E. Romero \and Daniela P\'erez}

\authorrunning{Gustavo E. Romero \and Daniela P\'erez} 
\institute{Gustavo E. Romero \and Daniela P\'erez
						\at Instituto Argentino de Radioastronom{\'\i}a, C.C.5, (1984)\\
              Villa Elisa, Bs. As., Argentina \\
              \email{romero@iar-conicet.gov.ar,\\
 							danielaperez@iar-conicet.gov.ar} 
							\and
							Gustavo E. Romero 
							\at Facultad de Ciencias Astron\'omicas y Geof{\'\i}sicas, UNLP, Paseo del Bosque s/n\\
							CP (1900), La Plata, Bs. As., Argentina}
\date{Received: 17 June 2011 / Accepted: 31 January 2012}

\maketitle

\begin{abstract}
We present a formal analysis of the Cosmological Argument in its two main forms: that due to Aquinas, and the revised version of the Kalam Cosmological Argument more recently advocated by William Lane Craig. We formulate these two arguments in such a way that each conclusion follows in first-order logic from the corresponding assumptions. Our analysis shows that the conclusion which follows for Aquinas is considerably weaker than what his aims demand. With formalizations that are logically valid in hand, we reinterpret the natural language versions of the premises and conclusions in terms of concepts of causality consistent with (and used in) recent work in cosmology done by physicists. In brief: the Kalam argument commits the fallacy of equivocation in a way that seems beyond repair; two of the premises adopted by Aquinas seem dubious when the terms `cause' and `causality' are interpreted in the context of contemporary empirical science. Thus, while there are no problems with whether the conclusions follow logically from their assumptions, the Kalam argument is not viable, and the Aquinas argument does not imply a caused origination of the universe. The assumptions of the latter are at best less than obvious relative to recent work in the sciences. We conclude with mention of a new argument that makes some positive modifications to an alternative variation on Aquinas by Le Poidevin, which nonetheless seems rather weak.
\keywords{Causality, cosmology, theology, semantics}
\end{abstract}

\section{Introduction}

\label{intro}

The so-called Cosmological Argument is a deductive argument for the existence of a cause, or a causal agent, of the universe. Contrary to the Ontological Argument, it is, like the Teleological Argument, an {\sl a posteriori} argument in the sense that the truth value of the premises should be known from our experience of the world \cite{Rowe98}.

The origin of the argument can be traced to Plato and Aristotle. In a modified form the Cosmological Argument can be found in the works of the Arabic theologian al-Ghazali in the eleventh century. Then, the argument was forcefully presented by Aquinas and Duns Scotus in the thirteenth century, modified again by Leibniz in the eighteenth century, and recently advocated by William Lane Craig \cite{Craig79}\cite{Craig80} and Richard Swinburne \cite{Swinburne79}. The Cosmological Argument came under serious assault in the eighteenth century, first by David Hume and then by Immanuel Kant. In the twentieth century, Bertrand Russell, John Mackie, Michael Martin, Adolf Gr\"unbaum, and many others have criticized different aspects of the argument. Contemporary discussions on the argument can be found, for instance, in Craig and Smith \cite{CraigSmith93}.

The Cosmological Argument, as it is clear from the variety of exposers, is actually a family of arguments that start from some alleged facts about the world and then, appealing to some general principle, proceed to establish the existence of a first cause, a prime mover, a necessary being, etc.

In this paper we shall be concerned with those forms of the Cosmological Argument that invoke causality, and hence are relevant to physical cosmological. Whether the concept of cause can be applied to the universe is an issue of considerable interest in current cosmology.

The purpose of this work is to analyze the structure of the two main forms of the argument, removing any vagueness inherent to the natural language. We shall study the formal structure of the argument. First, the argument will be translated to a purely syntactic formal language. We shall adopt a first order language with an infinite number of terms that allows for formulas formed by an infinite number of conjunctions. Descriptions of such language can be found, for instance, in Carnap \cite{Carnap58} and Rayo \cite{Rayo2004}. Second, the syntactic scheme will be interpreted in accordance with a realist ontology usually adopted in scientific practice. The correctness of the argument will then be pondered. An argument can be syntactically correct but its interpretation 
can be incorrect because of shifting in the meaning of some terms, vagueness, or other semantic pathologies.

We shall start our analysis with the second way of Aquinas, and then we shall proceed to analyze the modern interpretation of the Arabic form of the argument, as found in Craig \cite{Craig91}.

\section{The Cosmological Argument in Aquinas}

\label{sec:1}

In his  {\sl Summa Theologica}, Thomas Aquinas (1225-1274), writes:

\begin{quotation}

The second way is from the nature of the efficient cause. In the world of sense we find there is an order of efficient causes. There is no case known (neither is it, indeed, possible) in which a thing is found to be the efficient cause of itself; for so it would be prior to itself, which is impossible. Now in efficient causes it is not possible to go on to infinity, because in all efficient causes following in order, the first is the cause of the intermediate cause, and the intermediate is the cause of the ultimate cause, whether the intermediate cause be several, or only one. Now to take away the cause is to take away the effect. Therefore, if there be no first cause among efficient causes, there will be no ultimate, nor any intermediate cause. But if in efficient causes it is possible to go on to infinity, there will be no first efficient cause, neither will there be an ultimate effect, nor any intermediate efficient causes; all of which is plainly false. Therefore it is necessary to admit a first efficient cause, to which everyone gives the name of God\footnote{The Summa Theologica of St. Thomas Aquinas, Second and Revised Edition, 1920, 1a.2, 3. Literally translated by Fathers of the English Dominican Province. On-line Edition Copyright 2008 by Kevin Knight. }. 

\end{quotation}

The structure of the argument is quite simple. First we have a premise asserting a simple fact about the world: there are things with efficient causes. The efficient cause is the agent which brings something about. In modern terminology we can replace `efficient cause', simply by `cause', in the sense that will be specified below. After the first premise, we have two more general premises: one asserting that what is caused to exist is caused by something else, and then the final (and more controversial) premise with the denial of an infinite regress of causes\footnote{In Summa Theologica, Question 46, Article 2, Aquinas contends that it is an article of faith, as opposed to a matter of philosophical demonstration, that the world began. This, however, is not relevant for the issue under discussion here.}

In syllogistic form, the argument can be expressed as \cite{Rowe98}:

\begin{enumerate}

	\item{Some things exist, and their existence is caused.}

	\item{Whatever is caused to exist is caused to exist by something else.}

	\item{An infinite regress of causes resulting in the existence of a particular thing is impossible.}

	\item{Therefore, there is a first cause of existence.}

\end{enumerate}

In logical notation\footnote{As mentioned, we use a first-order formal language with infinite terms that allows for formulas formed by infinite conjunctions (See \cite{Carnap58} \cite{Rayo2004}).}:\\

$\:P_{1}.\:  (\exists x)\: {\mathfrak C}(x). $\\

$\:P_{2}.\:  (\forall x) \: \mathfrak{C}(x)\: \rightarrow \: (\exists y)\: {C}(x, y). $\\

$\:P_{3}.\: \sim (\exists x) \: \left[{C}(x, y_{1}) \wedge \bigwedge\limits_{i=1}^{\infty}{{C}(y_{i} , y_{i+1})}\right].  $\\

$\:D.\: (\exists z)\: \sim \mathfrak{C}(z). $\\

This argument is logically correct ($ P_{1} \wedge P_{2} \wedge P_{3}\vdash D $), {\em i.e.} there is at least one $z$ such that  $\sim \mathfrak{C}(z)$. The argument does not establish that there is one and only one $z$ such that  $\sim \mathfrak{C}(z)$. In other words, the domain of quantification of the bound variable $z$ in the consequence $D$ is restricted only to be non-empty. If we write, more explicitly:\\

$\:D.\: (\exists z)_{\mathfrak{D}}\: \sim \mathfrak{C}(z), $\\

\noindent then, we can only conclude that $\mathfrak{D}\neq\emptyset$. This is not what is stated by Rowe in the sense that ``there is a first cause of existence''. Such a statement cannot be derived from Aquinas's premises. The argument is compatible with any number of first causes.

In order to formulate an interpretation of the argument, we need to assign meanings to all extra logical symbols, that is to $\mathfrak{C}$ and $C$. We shall provide  definitions that are compatible with a formal, well-defined ontology of scientific practice and we shall then analyze whether the argument is still correct under such interpretation. But before, we shall briefly discuss the formal structure of the Kalam version of the argument.

\section{The Kalam Cosmological Argument}

The Kalam Cosmological Argument was first formulated by the medieval Muslim theologian al-Ghazali (1058-1111). He introduced a simple syllogism to
support the idea that the universe has a cause of its existence. This cause was attributed to the existence of a Creator. On the basis of philosophical arguments, al-Ghazali was able to avoid an infinite, temporal regress of past events, which was one of the main drawbacks in Aristotelian Cosmological Arguments. Persuaded that the universe begun to exist, al-Ghazali argued that it should have been causally originated as well.\


The traditional form of the Kalam Cosmological Argument is \cite{Craig91}:

\begin{enumerate}

	\item{Whatever begins to exist has a cause of its existence.}

	\item{The universe began to exist.}

	\item{Therefore, the universe has a cause of its existence.}

\end{enumerate}

In logical notation:\\

$\:P_{1}.\: (\forall x)\: B(x)\: \rightarrow (\exists y)\: {C}(x,\,y). $\\

$\:P_{2}.\: B(\sigma).$\\

$\:D.\: (\exists y)\: {C}(\sigma, \, y). $\\

This argument is logically correct ($ P_{1} \wedge P_{2}\vdash D $). Again, in order to make an interpretation of the argument, we need to assign meanings to all extra logical symbols, that is to $B$, $\sigma$ and ${C}$. In order to provide definitions that are compatible with the ontology of contemporary science and to analyze whether the argument remains correct under such assumptions, it is convenient to review first the basic ontological concepts we shall assume.

\section{Ontological background}

An ontology is a general theory about the nature of whatever exists.

The concept of individual can be considered the basic primitive concept of a formal ontological theory\footnote{We follow Bunge (1977) on the basics of our ontological views here. Other ontologies are possible, but for reasons discussed in Bunge (2010) we adopt an ontology of individuals and
properties that seems to be in good accordance with the existential assumptions usually made in physical science (see \cite{Bunge77} and \cite{Bunge10}.}. Individuals associate themselves with other individuals to yield new individuals. It follows that they satisfy a calculus, and that they are rigorously characterized only through the laws of such a calculus. These laws are set with the aim of reproducing the way real things associate. Specifically, it is postulated that every individual is an element of a set $s$ in such a way that the structure $\textsl{S}=\left\langle s, \circ, \square \right\rangle$ is a \textit{commutative monoid of idempotents}. This is a simple additive semi-group with neutral element.

In the structure \textsl{S}, $s$ is the set of all individuals, the element $\square \in s$ is a fiction called the null individual, and the binary operation $\circ$ is the association of individuals. Although \textsl{S} is a mathematical entity, the elements of $s$ are not, with the only exception of $\square$, which is a fiction introduced to form a calculus. The association of any element of $s$ with $\square$ yields the same element. The following definitions characterize the composition of individuals:

\begin{enumerate}

\item ${x} \in s $ is composed $ \Leftrightarrow  \left(\exists {y}, {z}\right)_{s} \left( {x} ={y} \circ {z} \right) $.

\item ${x} \in s $ is simple $ \Leftrightarrow \; \sim \left(\exists {y},{z}\right)_{s} \left({x} ={y} \circ {z} \right)$.

\item $ {x}\subset {y}\ \Leftrightarrow {x} \circ {y} = {y}\ $ (${x}$ is part of ${y}\ \Leftrightarrow {x} \circ {y} = {y} $).

\item $ \textsl{Comp}({x}) \equiv\{{y}\in s \;|\; {y}\subset {x}\}$ is the composition of ${x}$.\\

\end{enumerate}

An individual with its properties make up a thing $X$. Things can be represented by pairs of the form:
\[
X=<x,P(x)>,
\]	
where $P(x)$ is the collection of properties of the individual $x$. A material thing is an individual with material properties, {\em i.e.} properties that can change (see below) in some respect.

Things are distinguished from abstract individuals because they have a number of properties in addition to their capability of association. These 
properties can be \textit{intrinsic} \normalfont ($P_i$) or \textit{relational} \normalfont ($P_r$). The intrinsic properties are inherent and they are represented by predicates or unary applications, whereas relational properties depend upon more than a single thing and are represented by $n$-ary predicates, with $n \geq 1$. Examples of intrinsic properties are electric charge and rest mass, whereas velocity of macroscopic bodies and volume are relational properties\footnote{Velocity is an intrinsic property only in the case of photons and other bosons that move at the speed of light in any reference system.}.

If an individual is simple, its properties are called {\it elementary}, and the thing is {\it basic}. Examples of basic things in the Standard Model of particle physics are quarks and leptons, whereas examples of elementary properties are quark color and leptonic number, among others. 

The {\em state space} of a thing $X$ is a set of functions $S(X)$ from a domain of reference $M$ (a set that can be enumerable or non denumerable\footnote{In most physically interesting cases $M$ is a space-time continuum, see \cite{Bunge77}.} to the set of properties ${\cal P}_{X}$). Every function in $S(X)$ represents a property in ${\cal P}_{X}$. The set of the {\sl physically accessible} states of a thing $X$ is the {\em lawful state space} of $X$: $S_{\rm L}(X)$. The state of a thing is represented by a point in $S_{\rm L}(X)$. A change of a thing is represented by an ordered pair of states. Abstract things cannot change since they have only one state (their properties are fixed by definition). 

A {\em legal statement} is a restriction  upon the state functions of a given class of things. A {\em natural law} is a property of a class of material things represented by an empirically corroborated legal statement.

The {\em ontological history} $h(X)$ of a thing $X$ is a subset of $S_{\rm L}(X)$ defined by:
\[
h(X) = \{ \langle t, F(t) \rangle | t \in M\}, 
\] 
where $t$ is an	element of some auxiliary set $M$, and $F$ are the functions that represent the properties of $X$.

If a thing is affected by other things we can introduce the following definition:\\

$h(Y/X)$: ``history of the thing $Y$ in presence of the thing $X$''.\\

Let $h(X)$ and $h(Y)$ be the histories of the things $X$ and $Y$, respectively.\\

Then 
\[
h(Y/X) = \{ \langle t, H(t) \rangle |\: t \in M\}, 
\] 
where $ H\neq F $ is the total state function of $Y$ as affected by the existence of $X$, and $F$ is the total state function of $X$ in the absence of $Y$. The history of $Y$ in presence of $X$ is different from the history of $Y$ without $X$.\\

We can now introduce the notion of {\sl action}:\\

$ X \vartriangleright Y $: ``$X$ acts on $Y$''

\[
X \vartriangleright Y \stackrel{def}{=} h(Y/X)\neq h(Y). 
\] 
An {\sl event} is a change of a thing $X$, {\em i.e.} an ordered pair of states:
\[ 
(s_1, s_2 ) \in E_{\rm L}(X) = S_{\rm L}(X) \times S_{\rm L}(X). 
\]
The space $E_{\rm L}(X)$ is called the {\em event space} of $X$.\\

We shall use the concepts defined above to introduce a specific meaning for the term `causality'.

\section{Basic definitions: semantics}

\subsection{Aquinas's argument}\label{Aquinas}

The meaning of the premises in Aquinas's Second Way depends on the definition of `cause'.

In the context of the ontology sketched above, \textit{causality} is a relation between events, {\em i.e.} a relation between changes of states of
material things \cite{Bunge79}. It is {\sl not} a relation between things. Only the related concept of `action' is a relation between things. Specifically, we define:\\

$ \mathfrak{C}(x)$: ``an event in a thing $x$ is caused by some unspecified event $e^{x}_{x_{i}}$''.

$$ \mathfrak{C}(x)\stackrel{def}{=} (\exists x)(\exists e^{x}_{x_{i}}) \left[ e^{x}_{x_{i}}\in E_{\rm L}(x)\right] \Leftrightarrow  (\exists x_{i}) ( x_{i} \vartriangleright x).$$

${C}(x, y)$:``an event in a thing $x$ is caused by an event in a thing $y$''.

$$ {C}(x, y)\stackrel{def}{=} (\exists x)(\exists y)(\exists e^{x}_{y}) \left[ e^{x}_{y}\in E_{\rm L}(x) \right] \Leftrightarrow y \vartriangleright x. $$

In these definitions, the notation `$e^{x}_{y}$' indicates with the superscript the thing to whose event space belongs the event $e$, whereas the 
subscript denotes the thing that acted triggering the event. The implicit arguments of both $\mathfrak{C}$ and $C$ are events, not things. For simplicity in the notation we refer to the things that undergo the events. We shall also use hereafter lower case letters for variables that take values upon a domain of things.

Causation is a form of event generation. A crucial point is that a given event in the lawful event space $E_{\rm L}(x)$ is caused by an action of a thing $y$ iff the event happens only conditionally to the action, {\em i.e.}, it would not be the case of $e^{x}_{y}$ without an action of $y$ upon $x$. Time does not appear in this definition, allowing causal relations in space-times without a global time orientability and non-local causality.\\

\subsubsection{First premise}

The main problem with the usual interpretation of $P_{1}$ is that `existence' is not an event. What is an event is ``to begin to exist''. The event is the  emergence of the properties of a new thing $z$, its very first change. If we denote by $B(z)$ such a change (see Section \ref{Kalam-1} for a formal definition), we can reformulate $P_{1}$ as:\\

$\:P'_{1}.\:  (\exists x)\: (\exists y) \left[(y \vartriangleright x)\rightarrow  B(z) \right]. $\\

It is the action of $y$ upon $x$ what brings into existence the new thing $z$.

\subsubsection{Second premise}

The second premise can be reformulated in the following way:\\

$\:P'_{2}.\:  \sim (\exists x)\: \left[(x \vartriangleright x)\rightarrow
  B(x) \right]. $\\

Aquinas states that this is a self-evident premise: ``There is no case known (neither is it, indeed, possible) in which a thing is found to be the efficient cause of itself; for so it would be prior to itself, which is impossible.'' Romero and Torres \cite{RT} and Romero \cite{Romero04}, have shown, however, that it is possible to deny $\:P'_{2}$ without contradiction in space-times with non-trivial topology\footnote{Space-times with non-trivial topologies allow for timelike closed curves. An object moving along a closed curve can be self-existent in the following sense: it exists in a finite region of space-time, but it has neither beginning nor end. Nonetheless, every state of the object is causally related to a locally previous state. It has been suggested that the Universe itself might be such an object \cite{GottIIILi98}.}. Whether there are space-time regions with such a topology ({\em e.g.} in the interior of Kerr black holes or if there exist space-time wormholes) is an empirical issue, that should be set by empirical methods.

\subsubsection{Third premise}

The third premise is the denial of the possibility of an infinite number of causes or actions. In the light of the previous considerations on the 
meaning of `cause', we can rewrite this premise as:\\

$\:P'_{3}.\:  \sim (\exists z) \:  \left[\mathbb{B}(z, y_{1}) \wedge \bigwedge\limits_{i=1}^{\infty}{\mathbb{B}(y_{i} , y_{i+1})}\right], $\\

\noindent where

$$ \mathbb{B}(z,y) \stackrel{def}{=} (y \vartriangleright x)\rightarrow
  B(z). $$

The relation $\mathbb{B}$, according to this definition, implies generation with causality, whereas $B$ stands for unqualified generation. Under such interpretation, once again, the premise has an empirical content. It is not a self-evident principle. $\:P'_{3}$ assumes a linear topology for
events. In general (pseudo-Riemannian) space-times, closed time-like curves are possible. But even with a simple linear topology for the event space, whether such a space is finite or infinite is unclear.

\subsubsection{Reformulation of Aquinas's argument}

According to the previous considerations, Aquinas's argument can be formulated in the following form, which is both syntactically and semantically consistent:\\

$\:P'_{1}.\:  (\exists z)\: (\exists y) \: \mathbb{B}(z,y) $.\\

$\:P'_{2}.\:  \sim (\exists x)\: \left[(x \vartriangleright x)\rightarrow
  B(x) \right]. $\\

$\:P'_{3}.\:  \sim (\exists z) \:  \left[\mathbb{B}(z, y_{1}) \wedge \bigwedge\limits_{i=1}^{\infty}{\mathbb{B}(y_{i} , y_{i+1})}\right]. $\\

$\:D'.\: (\exists z)\: \sim B(z). $\\

Under the provided interpretation this argument satisfies  
$ P'_{1}\wedge P'_{2} \wedge P'_{3} \vdash D' $.

The truth value of  $\:P'_{2}$ and $\:P'_{3}$ is, however, controversial.

\subsection{Kalam argument}

\subsubsection{First premise}\label{Kalam-1}

The meaning of the first premise in the Kalam Cosmological Argument depends on the definitions of `to begin to exist' and `cause'.

We specify the meaning of these terms in scientific discourse as follows:
\\

\begin{itemize}

\item ``$x$ \textit{begins to exist}'' $\stackrel{def}{=}$ ``$x$ exists at time $t$ and there is a time interval $\Delta t\geq 0$ such that there  are no instants of time prior to $t-\Delta t$ at which $x$ exists'', see \cite{Romero04} for a detailed discussion.\\

In formal notation:\\

$ B(x) $: ``$ x $ \textit{begins to exists}''. \\

$B(x) \stackrel{def}{=} (\exists x)\;(\exists t_{*})  \; \left[x(t_{*})
,\, t_{*}\in \Re\right]\;\;\;\wedge \; \sim (\exists t) \; \left[x(t< t_{
*}-\Delta t), \;\;t\in \Re\right]. $\\

$ \Delta t \stackrel{def}{=} t_{2} - t_{1}, \; t_{2}\in \Re  ,\; t_{1}\
in \Re .$\\

\item Causality, as defined in the subsection \ref{Aquinas}, is a relation between events.

\end{itemize}

\subsubsection{Second premise}

The meaning of the second premise depends on the adopted definition of `universe'. According to the ontology assumed here \cite{Bunge77} \cite{PB}, the universe $(\sigma)$ is the composition of all things: 
\[ 
\sigma = [\Theta] \Leftrightarrow (x \subset \sigma \Leftrightarrow x \in \Theta). 
\] 
The set of all things is denoted by $\Theta$. The symbol $[\: ]$ represents the operation of composition. Composition is a relation among material things and should not be mistaken with `$\in$', which is a relation between elements and sets ({\em i.e.} abstract entities). In particular, the universe is not the set of all things. The universe is a physical entity with physical properties, like density, entropy and temperature.

\section{Does the universe have a cause?}

According to the definition of `cause', the universe should not include at least one thing in order to have a cause. This condition contradicts the very definition of universe. Then, the usual interpretation of the Kalam Cosmological Argument is not sound, since the meaning of ${C}$ is different in the premise $P_{1}$ and in the consequence $D$. It follows that the argument is not valid under the present interpretation. A new one should be provided or the argument abandoned.

Aquinas's form of the argument, although formally correct, says nothing about the cause of the universe. His argument, 
however, can be easily cast into a form with explicit advocacy for a cause of the universe. In the version presented by Robin Le Poidevin, the argument goes as follows \cite{Pondevin96}:

\begin{enumerate}

	\item{Anything that exists has a cause of its existence.}

	\item{Nothing can be the cause of its own existence.}

	\item{The universe exists.}

	\item{Therefore, the universe has a cause of its existence which lies outside the universe.}

\end{enumerate}

This form is incorrect under the above given definition of universe, since the expression ``outside the universe'' makes no sense. It
can be modified, nonetheless, introducing some restriction in the premises, regarding the spatio-temporal nature of physical existence. For instance:

\begin{enumerate}

	\item{Anything that exists in space-time has a cause of its existence.}

	\item{Nothing in space-time can be the cause of its own existence.}

	\item{The universe exists spatio-temporally.}

	\item{Therefore, the universe has a cause of its existence outside space
-time.}

\end{enumerate}

This form introduces some new terms. Space-time is the physical sum or composition of all events of all things. To say that ``something exists in space-time'' is just a way of saying that something has spatio-temporal relations with other things \cite{PB}. Since events are ordered pairs of states of things, the existence of space-time requires the existence of interacting things. Whether a cause can be defined outside space-time seems highly dubious \cite{G89} \cite{G90} \cite{G91} \cite{G00}.

\section{The relation between physical and philosophical cosmology}

According to the preceding analysis, physical cosmology has precedence over philosophical cosmology. In fact, a given interpretation of a cosmological argument can be at odds with observational facts. For instance, some of the premises of the Cosmological Argument can be interpreted as expressing a necessary truth, when actually the validity of the premises might depend on the way things are in the universe. An example of this is the assertion by Aquinas that ``There is no case known (neither is it, indeed, possible) in which a thing is found to be the efficient cause of itself; for so it would be prior to itself, which is impossible". This statement is far from being a necessary truth (i.e. one which negation implies contradiction). The truth value of the statement, as already mentioned in
 Section 5.1.2. and discussed in the references cited therein, depends on the structure of the space-time. In space-times with time-like closed curves is possible, in principle, to have objects that never began to exist but whose changes of state are always causally linked to previous changes. This causal relation is always locally future-directed, but never globally, because this general-relativity admissible space-times are not globally hyperbolic (and hence they present Cauchy horizons). Whereas there exist in the universe matter with energy-momentum distributions capable of shaping space-time curvature to make such self-existent objects possible, is a matter of fact and not of philosophical analysis.

By other hand, philosophical cosmology can shed light on our assignation of meanings to different models of syntactic arguments, allowing to generate coherence tests for such arguments. The value of the conclusions will depend on the value of the premises, which sometimes ought to be assigned in accordance with observations or experiments.

Summing up, we can say that in philosophical cosmology speculative metaphysics meets experiment, for the benefit of our understanding of the world.

\section{Final remarks}

Both causation and space-time are intimately related to change, {\em i.e.} to events, ordered pairs of states of things. It is very difficult, if 
not impossible, to make sense of the concept of causation outside space-time. This in no way means that the universe cannot have an {\sl explanation}. Causation is only one possible form of event generation, but not the only one. Decay and spontaneous symmetry breaking are examples of event generation without causation. It is very likely that an explanation of the origination of the universe should involve a quantum theory of gravity. In such a theory, space-time at the Planck-scale would not have a causal structure since it would not be represented by a continuous and differentiable manifold endowed with an affine connection. Aquinas's argument correctly suggests that some happenings in the world are causeless. Likely, this reasoning is valid for the universe as a whole, since cause and effect are concepts that make sense only {\sl within} the universe, because they are unavoidably linked to continuum space-time. We conclude that what
is the ultimate explanation for the universe cannot be established from our present knowledge of the world.








\bibliographystyle{spbasic}      




\begin{thebibliography}{}

\bibitem{Rowe98}

Rowe, W.L. (1998). {\em The Cosmological Argument}. Princeton: Princeton University Press.


\bibitem{Craig79}

Craig, W.L. (1979). {\em The Kalam Cosmological Argument}. New York: Harper \& Row.



\bibitem{Craig80}

Craig, W.L. (1980). {\em The Cosmological Argument from Plato to Leibniz}. London: The Macmillan Press.

\bibitem{Swinburne79}

Swinburne, R. (1979). {\em The existence of God}. Oxford: Clarendon Press.

\bibitem{CraigSmith93}

Craig, W.L. \& Smith, Q. (1993). {\em Theism, atheism, and Big Bang cosmology}. Oxford: Clarendon Press.

\bibitem{Carnap58}

Carnap, R.(1958). {\em Introduction to symbolic logic and its applications}. New York: Dover Publications.

\bibitem{Rayo2004}

Rayo, A. (2004). Formalizaci\'on y lenguaje Ordinario. In R. Orayen \& A. Moretti (Eds.), Filosof\'{\i}a de la L\'ogica (pp. 17-42). Madrid: Trotta.

\bibitem{Craig91}

Craig, W.L. (1991). The existence of God and the beginning of the universe. {\em Truth}, 3, 85-96.

\bibitem{Bunge77}

Bunge, M. (1977). {\em Treatise of basic philosophy. Ontology I: the furniture of the world}. Dordrecht: Reidel.

\bibitem{Bunge10}

Bunge, M. (2010). {\em Matter and mind: a philosophical inquiry}. Dordrecht: Springer.


\bibitem{Bunge79}

Bunge, M. (1979). {\em Causality and modern science}. New York: Dover.


\bibitem{RT}

Romero, G.E., \& Torres, D.F. (2001). Self-existing objects and auto-generated Information in chronology-violating space-times. {\em Modern Physics Letters A} 16, 1213-1222.



\bibitem{Romero04}

Romero, G.E. (2004). God, causality, and the creation of the universe. {\em Invenio}, 13, 11-20.



\bibitem{GottIIILi98}

Gott III, J.R. \& Li, L.X. (1998). Can the universe create itself?. {\em Phys. Rev. D}, 58, 023501.  


\bibitem{PB}

Perez-Bergliaffa, S.E, Romero, G.E. \& Vucetich, H. (1998). Toward an axiomatic pregeometry of space-time. {\em International Journal of Theoretical Physics}, 37, 2281-2298.


\bibitem {Pondevin96}

Le Pondevin, R. (1996). {\em Arguing for atheism}. London and New York: Routledge.




\bibitem{G89}

Gr\"unbaum, A. (1989). The Pseudo-problem of creation in physical cosmology. {\em Philosophy of Science}, 56, 373-394.



\bibitem{G90}

Gr\"unbaum, A. (1990). {Pseudo-creation of the Big Bang}. {\em Nature}, 344, 821-822.



\bibitem{G91}

Gr\"unbaum, A. (1991). Creation as a pseudo-explanation in current physical cosmology. {\em Erkenntnis}, 35, 233-254.



\bibitem{G00}

Gr\"unbaum, A. (2000). A new critique of theological interpretations of physical cosmology. {\em British Journal for the Philosophy of Science}, 51, 1-43.

\end{thebibliography}


\end{document}